\documentclass[%
 reprint,
 amsmath,amssymb,
 aps,
floatfix,
]{revtex4-2}

\usepackage{graphicx}% Include figure files
\usepackage{dcolumn}% Align table columns on decimal point
\usepackage{bm}% bold math
\usepackage{placeins}
\usepackage{xcolor}
	\usepackage{color}
	\usepackage{soul}

\begin{document}
\preprint{APS/123-QED}

%usefull commands:
	\newcommand{\rvec}{\mathbf{r}}
	\newcommand{\Rvec}{\mathbf{R}}
	\newcommand{\kvec}{\mathbf{k}}
	\newcommand{\Gvec}{\mathbf{G}}
	\newcommand{\avec}{\textbf{a}}
	\newcommand{\bvec}{\textbf{b}}
	\newcommand{\cvec}{\textbf{c}}
	\newcommand{\pvec}{\textbf{p}}
	\newcommand{\pivec}{\pmb{\pi}}
	\newcommand{\sigmavec}{\pmb{\sigma}}
	\newcommand{\ksplit}{k_\text{split}}
	\newcommand{\Esplit}{E_\text{split}}
	\newcommand{\Egap}{E_\text{gap}}
	\newcommand{\dmax}{d_\text{max}}
	\newcommand{\exciting}{\texttt{exciting}}
	\newcommand{\splitparam}{eV$\cdot$Å}
	
	\newcommand{\mapi}{$\text{MA}\text{Pb}\text{I}_\text{3}$}
	\newcommand{\bulk}{$\text{Cs}_\text{2}\text{Pb}\text{I}_\text{3}$}
	\newcommand{\none}{$\text{Cs}_\text{2}\text{Pb}\text{I}_\text{4}$}
	
	\newcommand{\angstr}{\mathring{A}}
\newcommand{\blue}[1]{\color{blue}#1}
\newcommand{\red}[1]{\color{red}#1}
\newcommand{\magenta}[1]{\color{magenta}#1}
\newcommand{\green}[1]{\color{green}#1}
\newcommand{\cyan}[1]{\color{cyan}#1}

\preprint{APS/123-QED}

\title{{Rashba and Dresselhaus effects in 2D Pb-I-based perovskites}}

\author{Benedikt Maurer}
\author{Christian Vorwerk}
\author{Claudia Draxl}
\affiliation{%
 Institut f\"ur Physik and IRIS Adlershof, Humboldt-Universit\"at zu Berlin, Berlin Germany
}%

\date{\today}

\begin{abstract}
Bulk hybride halide perovskites are governed by significant Rashba and Dresselhaus splitting. This indicates that such effects will not only affect their optoelectronic properties but also those of their two dimensional layered relatives. This work aims at understanding how different ways of symmetry breaking influence these effects in those materials. For this purpose, model structures are adopted where the organic compounds are replaced by Cs atoms. Disregarding possible distortions in the inorganic layers,  results in structures with composition Cs$_{n+1}$Pb$_n$I$_{3n+1}$. Using the all-electron full-potential density-functional-theory code \texttt{exciting}, the impact of atomic displacement on the band structure is systematically studied for $n=1$, 2, 3 and $\infty$. The displacement patterns that yield Rashba or Dresselhaus splitting are identified, and the amount of the splitting is determined as a function of displacement. Furthermore, the spin textures in the electronic states around the band gap are analyzed to differentiate between Rashba and Dresselhaus effects. This study reveals in-plane Pb displacements as the origin of the strongest effects.
\end{abstract}

\maketitle

\section{Introduction}
The identification of candidate materials for efficient and environmentally friendly devices in optoelectronics and photovoltaics requires an accurate understanding of their electronic structure. 
In  the last decade, hybrid organic-inorganic halide perovskites (HaP) have become of particular interest due to their impressive performance as photovoltaic materials combined with low-cost production \cite{Kojima2009,Zhou2014,Snaith2013, Green2014, Gratzel2014, Berry2015, Brenner2016, Nair2020, Liang2020, Dequilettes2019, Kim2020a, Jena2019, Kim2020}.
In particular, the 2D HaPs have been in the focus of recent extensive research \cite{Mitzi1994, Mitzi1995, Kagan1999, Smith2014, Cao2015, Tsai2016, Kim2020b, Blancon2020, Lian2020}. They consist of atomically thin layers of HaPs, often referred to as inorganic layers  that are separated by organic molecules. The thickness of these inorganic layers varies by the number of metal-halide octahedra stacked on each other (labeled $n$). Due to the alternation of organic and inorganic layers, 2D HaPs form natural quantum wells. 
They have been in the focus of recent extensive research, because of their promising potential for applications in optoelectronics, microelectronics, and photovoltaics \cite{Mitzi1994, Mitzi1995, Kagan1999, Tsai2016, Cao2015}. Additionally, they show better environmental stability than 3D HaPs \cite{Smith2014, Lian2020, Kim2020b}, which is a great advantage for the use in photovoltaics. 

2D and 3D HaPs contain heavy elements such as Pb and I and show a great structural diversity, often with lack of inversion symmetry. 
The presence of heavy atoms implies that spin orbit coupling (SOC) has a great influence on the electronic structure. 
Combined with the lack of inversion symmetry, SOC can lift the spin degeneracy around the band gap.
This band splitting was first discovered for zinc-blende structures by Dresselhaus \cite{Dresselhaus1955} and later for wurzite structures  by Rashba and Bychov  \cite{Bychkov1984}. In general, structure inversion asymmetry (SIA) leads to Rashba, and bulk inversion asymmetry (BIA) to Dresselhaus effects \cite{Winkler2003}. Both effects have been studied extensively for many materials. They have been observed in heterostructures, \cite{Lommer1988, Nitta1997, Rechcinski2021}, quantum wells \cite{Schultz1996, Engels1997, Balocchi2013, Abid2012, Wang2014, Rastegar-Sedehi2021}, bulk materials {\cite{Ishizaka2011, Eremeev2012, Ideue2014}}, heavy-atom systems, and alloy surfaces \cite{Dil2008, Eremeev2012, LaShell1996, Koroteev2004, Ast2007, Takayama2012, Bianchi2012, Eremeev2013, Wang2015}, as well as in nanowires \cite{Banerjee2011,Liang2012,Zhang2015}. 
Recently, these effects led to the development of topological superconductors for quantum-information processing through Majorana fermions \cite{Sau2010, Alicea2010}. 
In addition, SOC effects also enable the control of spin-dependent band structures
which opens the field for spintronic applications \cite{Gregg2002, Jansen2003, Zutic2004, Jansen2012}.

Rashba and Dresselhaus effects have been studied in 3D HaPs in the last years by many authors \cite{Kim2014, Zheng2015, Kepenekian2015a, Etienne2016, Niesner2016, Mosconi2017a, McKechnie2018}. For 2D HaPs, Rashba splitting has been reported in several materials \cite{Kepenekian2015a, Zhai2017, Yin2018, Todd2019, Park2019}, for example, in phenyl-ethyl-amonium-lead-iodide (PEPI; n=1) %(C$_6$H$_5$C$_2$H$_4$NH$_3$)$_2$PbI$_4$ 
and in phenyl-ethyl-amonium-methyl-amonium-lead-iodide (PEMAPI; $n=2$) %(C$_5$H$_5$C$_2$H$_4$NH$_3$)$_2$(CH$_3$NH$_3$)Pb$_2$I$_7$
\cite{Yin2018}. In contrast to the results in Ref. \cite{Zhai2017}, Rashba splitting was not confirmed in PEPI with $n=1$ \cite{Yin2018}.
In these studies Rashba and Dresselhaus effects are not explicitly distinguished. Even the presence of Rashba splitting in PEPI with $n=1$ is reported controversially, since Zhai and coworkers find Rashba splitting for this material \cite{Zhai2017} in contrast to Yin and coworkers \cite{Yin2018}. 
%These studies refer only to the Rashba effect since they do not distinguish between Rashba and Dresselhaus effect.
SOC effects in 2D HaPs with $n>2$  have not been addressed so far. 
As such, there is a need to establish a possible connection between layer thickness and the strength of SO splitting.
The structural complexity of the 2D HaPs additionally calls for asking to which extent a specific way of breaking inversion symmetry leads to Rashba and/or Dresselhaus splitting. 

To answer these questions, we introduce a 2D model perovskite with inversion-symmetry, and systematically investigate from first principles the influence of various ways of symmetry breaking on the electronic structure. 
We study the influence of layer thickness on the Rashba and/or Dresselhaus effect in this model system for $n=1, 2, 3$ and $\infty$.

\section{Methods}
\subsection{Modeling Rashba and Dresselhaus effects}
\begin{figure}
\includegraphics{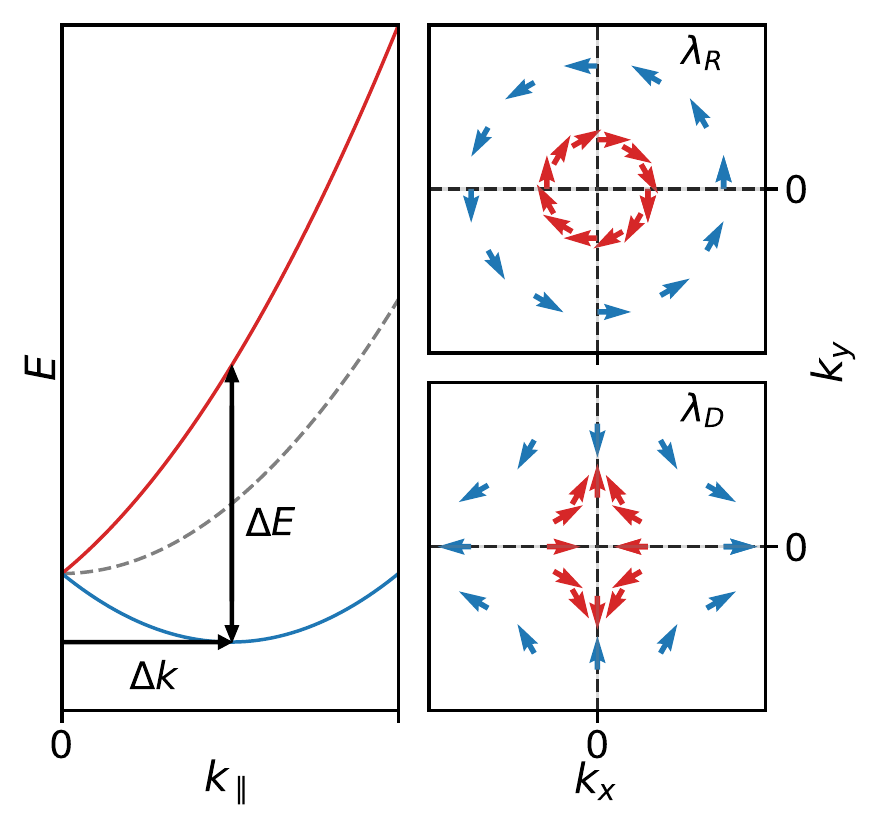}% Here is how to import EPS art
\caption{\label{fig:2deg}Left panel: Band structure of the free-electron gas with broken inversion symmetry with (red and blue lines), and  without (grey dashed line) SOC splitting. Right panel: Characteristic spin texture around the band gap for the Rashba effect (upper panel) and for the Dresselhaus effect (lower panel).}
\end{figure}

Spin-orbit coupling corresponds to the interaction of the spin with the orbital degrees of freedom \cite{Thomas1926}. From a non-relativistic approximation of the Dirac equation, SOC enters the Hamiltonian through the Pauli term \cite{Sakurai1967}:
\begin{equation}\label{HSOC}
	H_{SOC} = \frac{1}{4 c^2}  \left( \sigmavec \times \nabla V \right) \cdot \pvec,
\end{equation}
where $c$ is the velocity of light, $\pvec$ the electron momentum, $\sigmavec = (\sigma_x, \sigma_y, \sigma_z)$ the Pauli matrices, and $V$ the electrostatic potential in which the electron moves, induced by the nuclei.
Due to the linear dependence of SOC on $\nabla V$ and $\pvec$, it becomes large for electrons that move with large momentum in a potential with a considerable potential gradient perpendicular to the momentum. 
This is the case for electrons in the potential of heavy nuclei, such as Pb or I \cite{Even2012,Even2013}.

The SOC Hamiltonian is invariant under time reversal, thus Kramer' s degeneracy rules yield for the single-particle band energies:
\begin{equation}
	\epsilon_{n\uparrow}(\kvec) = \epsilon_{n\downarrow}(-\kvec)\text{ and }\epsilon_{n\downarrow}(\kvec) = \epsilon_{n\uparrow}(-\kvec).
\end{equation}
If inversion symmetry is present, another set of degeneracy rules applies:
\begin{equation}
	\epsilon_{n\uparrow}(\kvec) = \epsilon_{n\uparrow}(-\kvec)\text{ and }\epsilon_{n\downarrow}(\kvec) = \epsilon_{n\downarrow}(-\kvec).
\end{equation}
Combining both symmetries leads to double spin degeneracy of the bands within the entire Brillouin zone (BZ):
\begin{equation}\label{doubledeg}
	\epsilon_{n\uparrow}(\kvec) = \epsilon_{n\downarrow}(\kvec),
\end{equation}
Broken inversion symmetry can lift it along some path of the BZ. 
If this occurs at the band gap, this effect is known as the Rashba or the Dresselhaus effect \cite{Dresselhaus1955, Bychkov1984}. 
In this case, the bands related to the two spins are displaced against each other and thus the position of the band gap may be shifted. In the vicinity of the band gap without Rashba or Dresselhaus effect, the presence of SOC acts in first order like an effective magnetic field $\mathbf{\Omega}(\kvec)$ \cite{Winkler2006}. 
Thus a good approximation to the energy splitting is a Zeeman term:
\begin{equation}
	\Delta E(\kvec) = \frac 1 2 \sigmavec\cdot\mathbf{\Omega}(\kvec)= \frac 1 2 \sigmavec\cdot(\boldsymbol{\lambda}_R(\kvec) + \boldsymbol{\lambda}_D(\kvec)).
\end{equation}
The effective magnetic field $\mathbf{\Omega}(\kvec)$ can be approximated within $\kvec\cdot\pvec$ theory \cite{Winkler2003}, which enables us to separate the contributions from the Rashba effect and Dresselhaus effect, termed $\boldsymbol{\lambda}_R(\kvec)$ and $\boldsymbol{\lambda}_D(\kvec)$, respectively.

Before we consider them in the 2D perovskites with broken inversion symmetry, we first look at the 2D free electron gas (FEG) with $C_{2\nu}$ point symmetry, where analytical solutions are available \cite{Vajna2012, Kepenekian2015a}. We choose the Cartesian coordinate system such that the $C_2$ axis corresponds to the $z$ axis, and the 2D FEG is placed in the $(x, y)$ plane. The momentum perpendicular to the plane is then given by $\kvec_\perp=\kvec_z$ and the in-plane component by $\kvec_\parallel=(k_x, k_y, 0)$. Up to linear terms in $\kvec_{\parallel}$, the SOC can be approximated by the Rashba-Dresselhaus Hamiltonian \cite{Vajna2012, Kepenekian2015a}:
\begin{equation}
	H_{RD} = \lambda_R(k_x\sigma_y-k_y\sigma_x) + \lambda_D(k_x\sigma_x-k_y\sigma_y),
\end{equation}
where $\lambda_R$ and $\lambda_D$ are scalar parameters related to the two contributions.
Applying this term to the FEG, lifts the degeneracy of $E_0=\frac{1}{2} |\kvec_\parallel|^2$ (see Fig.~\ref{fig:2deg}) and leads to solutions for the inner ($+$) and the outer branch ($-$) (see Fig.~\ref{fig:2deg}):
\begin{equation}
	E_{RD\pm}  =  \frac{1}{2}|\kvec_{\parallel}|^2 \pm  \sqrt{(\lambda_D^2+\lambda_R^2)(k_x^2+k_y^2)-4\lambda_R\lambda_Dk_xk_y},
\end{equation}
\begin{equation}\label{eigvec_xy}
	\Psi_{RD\pm}(\kvec_{\parallel}) = \frac{e^{i\kvec_{\parallel}\cdot\rvec}}{2\pi}\frac 1{\sqrt{2}}
	\begin{pmatrix} \mp \frac{-\lambda_D(k_x+ik_y) + i\lambda_R(k_x-ik_y)}
	{\sqrt{(\lambda_D^2+\lambda_R^2)(k_x^2+k_y^2)-4\lambda_R\lambda_Dk_xk_y}} \\1
	\end{pmatrix}.
\end{equation}
For either pure Rashba or Dresselhaus effect ($\lambda_{D/R}=0$), the energy splitting is given by:
\begin{equation}\label{splittot}
    \begin{split}
	    \Delta E(\kvec_{\parallel}) &= E_{RD+}(\kvec_{\parallel})-E_{RD-}(\kvec_{\parallel})\\
	    &=2\lambda_{R/D} k_{\parallel},
	\end{split}
\end{equation}
where $k_{\parallel}=\sqrt{k_x^2+k_y^2}$.
The respective parameter can now be directly determined by
\begin{equation}
    \lambda_{R/D}=\frac{\Delta E(\kvec_{\parallel})}{2\sqrt{k_x^2+k_y^2}}.
\end{equation}
In the case of a mixed effect ({\it e.g.}, $\lambda_R\neq 0$ and $\lambda_D\neq 0$) the splitting parameters cannot be determined from the band structure alone. Rashba and Dresselhaus effect can be discriminated by considering the spin texture $\langle \sigmavec \rangle_{RD\pm} = \langle \Psi_{RD\pm}| \sigmavec | \Psi_{RD\pm} \rangle$. Symmetry suggests to transform the system into polar coordinates, with $\kvec=k_\parallel(\cos\theta, \sin\theta, 0)$ thus the eigenvectors (Eq.~\ref{eigvec_xy}) transform to
\begin{equation}
	\Psi_{RD\pm}(\kvec_{\parallel}) = \frac{e^{i\kvec_{\parallel}\cdot\rvec}}{2\pi}\frac 1{\sqrt{2}}
	\begin{pmatrix} \mp \frac{-\lambda_De^{i\theta} + i\lambda_Re^{-i\theta} }{\sqrt{(\lambda_D^2+\lambda_R^2)-2\lambda_R\lambda_D\sin(2\theta)}} \\1
	\end{pmatrix}.
\end{equation}
The spin texture for the general case is then given by
\begin{equation}\label{spintext}
	\begin{split}
		\langle \sigmavec \rangle_{RD\pm}\propto
		%\pm  \frac{1}{\sqrt{(\lambda_D^2+\lambda_R^2)-2\lambda_R\lambda_D\sin(2\theta)}}
		\begin{pmatrix}
		-\lambda_R\sin(\theta) + \lambda_D\cos(\theta)  \\ \lambda_R\cos(\theta) -\lambda_D\sin(\theta)   \\ 0
		\end{pmatrix}.
	\end{split}		
\end{equation}
\begin{figure}[t!]
\includegraphics{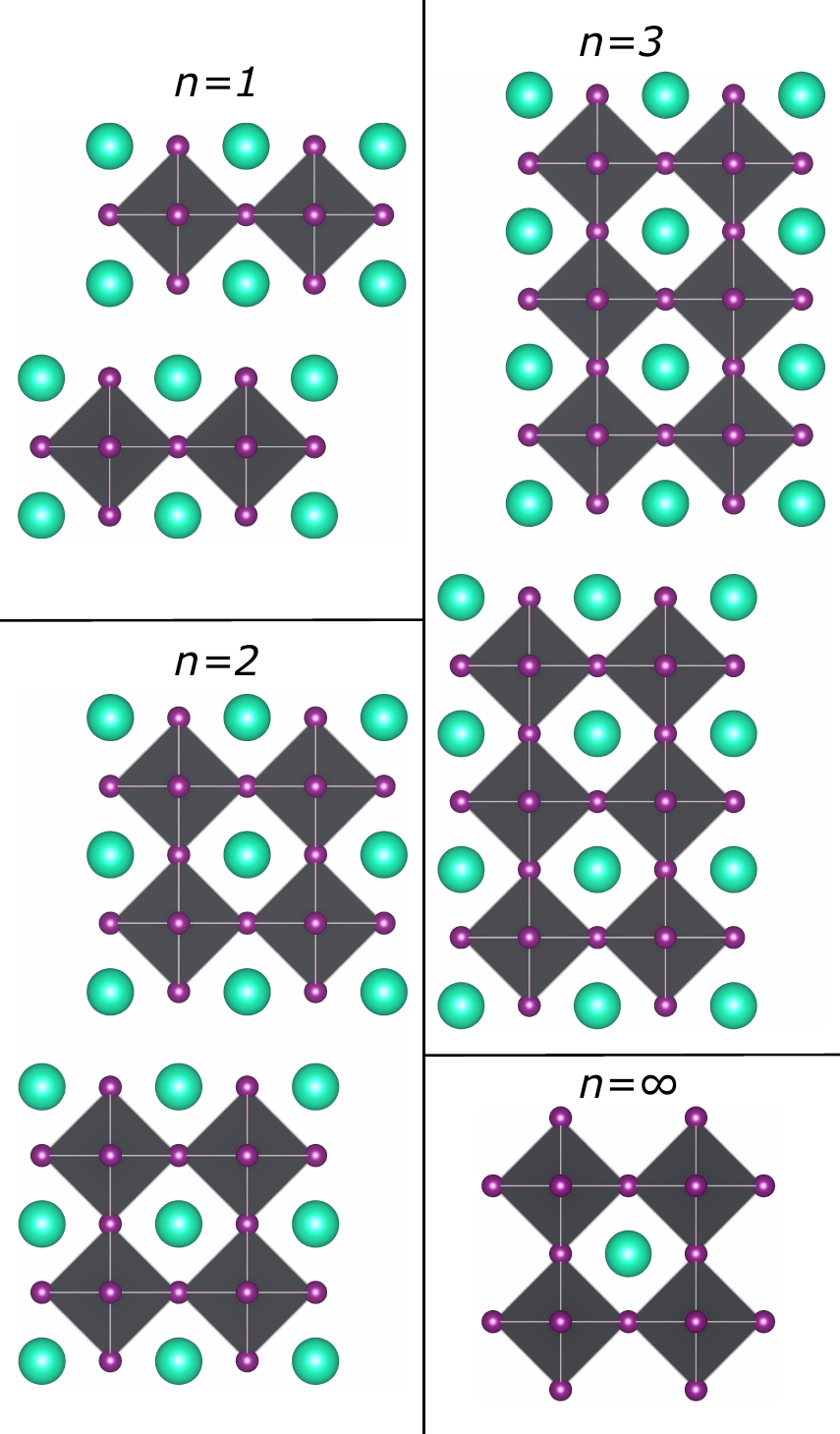}% Here is how to import EPS art
\caption{\label{fig:modelpero}Structures investigated in this work, represented by the chemical formula Cs$_{n+1}$Pb$_n$I$_{3n+1}$ for $n=1,2,3$ and $\infty$}.\label{2DFEGspino}
\end{figure}
For the pure Rashba effect ($\lambda_D=0$) the polar-coordinate angle of the spin direction and the corresponding {\bf k}-vector have a constant difference of $\pi/2$, leading to a circular spin texture (Fig. \ref{fig:2deg}, upper right panel). In case of the pure Dresselhaus effect ($\lambda_R=0$), the polar-coordinate angle of the spin direction runs against the angle of $\kvec_{\parallel}$ which results in a hyperbolic spin texture (Fig. \ref{fig:2deg}, lower right panel). Once the spin texture is determined for a given system, the relation $\lambda_R / \lambda_D$ can be obtained from Eq.~\ref{2DFEGspino} and the total strength ($\sqrt{\lambda_D^2 + \lambda_R^2}$) from the band splitting at the band gap.
\begin{figure*}
    \includegraphics{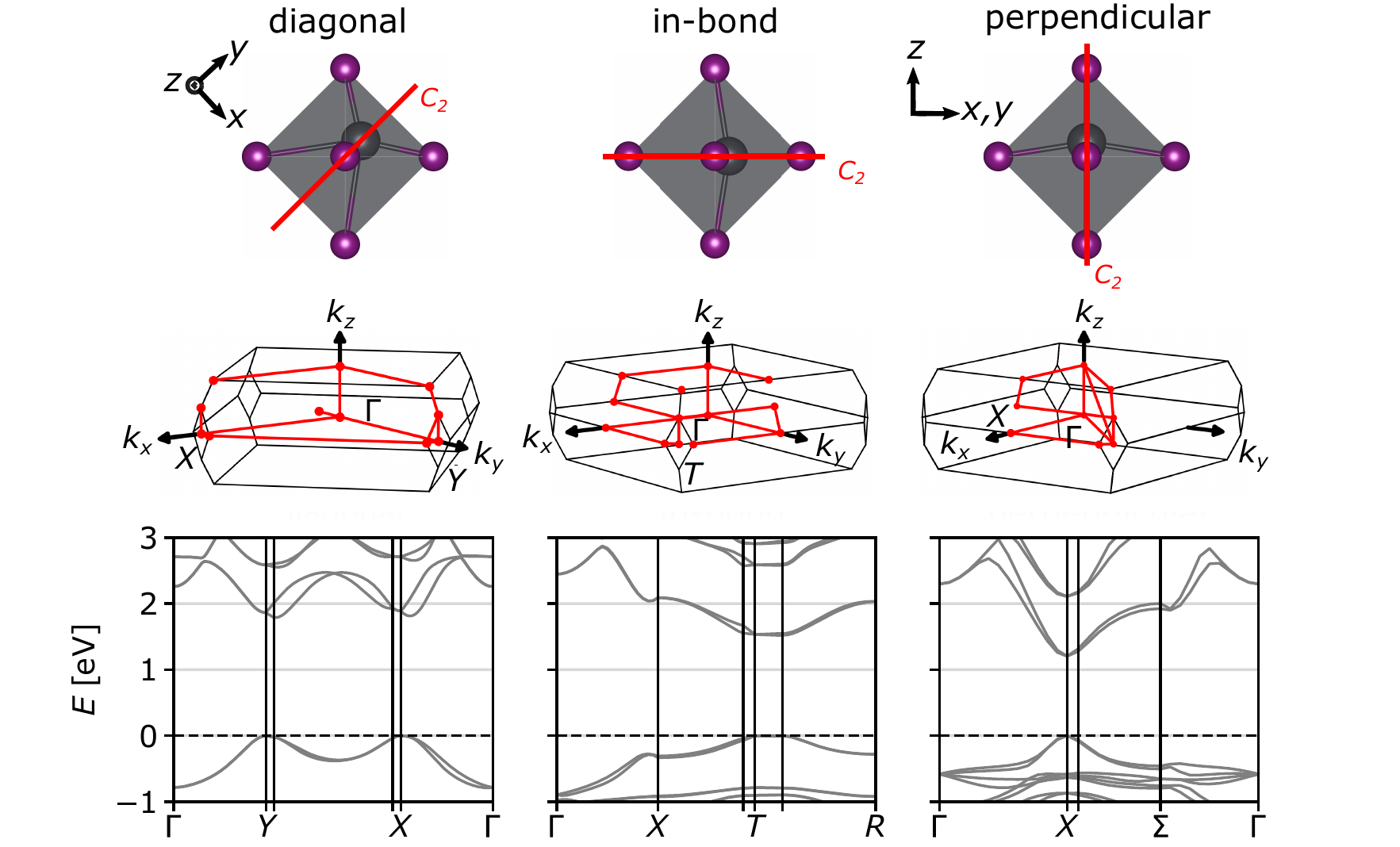}
    \caption{\label{fig:dispdiag} PbI$_6$ octahedra of Cs$_{2}$PbI$_{4}$ with diagonal, in-bond, and perpendicular displacements of the Pb atoms (top), corresponding Brioullin zones with the high symmetry paths in red (middle), and  band structures along selected high-symmetry paths (bottom).} 
 \end{figure*}

\subsection{Model perovskites}
2D hybrid organic-inorganic perovskites are built by alternating layers of the bulk perovskite structure, which, in turn, are composed of Pb-I octahedra and organic molecules (such as methyl-ammonium), placed in the cavities between the octahedra, and layers of larger organic molecules. 
Commonly, the perovskite layers are denoted as the inorganic layers and the dividing molecules as the organic layers.
A large variety of such structures exists.
On the one hand, this is due to the variable thickness of the inorganic layers, defined by the number of octahedral layers $n$ where for $n=\infty$, 3D perovskites are obtained.
On the other hand, many different organic molecules can be used as the organic layer. The size and structure of these molecules influence the distortion inside the inorganic layers and thus the symmetry of the material \cite{Stoumpos2016}. This structural variety makes the investigation of the nature of the SOC effects a difficult task.

In this study, we start with highly symmetric structures of the type Cs$_{n+1}$Pb$_n$I$_{3n+1}$ for $n=1$, 2, 3, and $\infty$ (see Fig. \ref{fig:modelpero}). 
In these structures, the organic molecules are replaced with Cs atoms, and all distortions are removed from the octahedral lattice.
These structures are body-centered tetragonal with the symmetry group \textit{I4mmm} for $n=1$, 2, and 3  and primitive cubic for $n=\infty$.
As all of them have inversion symmetry, neither Rashba nor Dresselhaus effect occurs.
As several works report octahedral distortions as the origin of broken inversion symmetry in 2D HaPs \cite{Stoumpos2016, Zhai2017, Yin2018}, we aim at modeling distortions that break inversion symmetry.
To this extent, we introduce uniform Pb displacements in either \textit{diagonal} direction with respect to the Pb-I bonds that lie  in-plane of the layers (Fig. \ref{fig:dispdiag}, upper left panel), in direction of the in-plane Pb-I bonds (\textit{in-bond}; Fig. \ref{fig:dispdiag}, upper center panel), or \textit{perpendicular} to the plane (Fig. \ref{fig:dispdiag}, upper right panel).
The energy splitting ($\Delta E$) is extracted from the band structure around the band gap and the splitting parameters ($\lambda_D$ and $\lambda_R$) from the spin texture with help of Eq.~\ref{spintext}.
This approach allows us to separate the impact of lattice distortion and layer thickness on the electronic structure and furthermore, differentiate between Rashba and Dresselhaus effect.

\subsection{Computational details}

All calculations are carried out with the all-electron full-potential package \texttt{exciting} \cite{Gulans2014}. The electronic ground state is calculated within density-functional theory (DFT), using the GGA-PBE xc functional \cite{Perdew2008}.
The  $k$-grid and the plane wave cutoff $|\mathbf{G}+\mathbf{k}|_{max}$ used in the calculations that are considered to be well converged are displayed in Table \ref{tab:parameters}.

\begin{table}
\caption{$k$-grids and plane wave cutoffs used in the calculations.}
\label{tab:parameters}
\begin{tabular}{ |c|c|c|c|c| } 
\hline
$n$ & 1 & 2 & 3 & $\infty$ \\
\hline
$k$-grid & $7 \times 7 \times 7$ & $6 \times 6 \times 6$ & $6 \times 6 \times 6$ & $7 \times 7 \times 7$ \\ 
$|\mathbf{G}+\mathbf{k}|_{max}$ & 2.8 a.u. & 4.4 a.u. & 3.2 a.u. & 3.6 a.u. \\
\hline
\end{tabular}
\end{table}

The spin texture is evaluated around the Kohn-Sham band gap. This is done for Cs$_{2}$Pb$_1$I$_{4}$ with ten equidistant Pb displacements in {\it diagonal} direction in the range of $[0.0\text{ Å}, 0.56\text{ Å}]$, in {\it in-bond} direction in the range of $[0.00\text{ Å}, 0.50\text{ Å}]$, and in {\it perpendicular} direction in the range of $[0.00\text{ Å}, 0.50\text{ Å}]$. For Cs$_{n+1}$Pb$_n$I$_{3n+1}$ with $n=2$, 3, and $\infty$, {\it diagonal} Pb displacements are considered in the range of $[0.00\text{ Å}, 0.56\text{ Å}]$. The symmetries of the investigated structures are analyzed with the \textit{Python Materials Genomics} package \textit{(pymatgen)} \cite{Ong2013}.
All input and output files, can be found in the NOMAD Repsitory: https://dx.doi.org/10.17172/NOMAD/2021.11.08-1.

\section{Results}
{Origin of Rashba and Dresselhaus effects}
\begin{figure}[t!]
\includegraphics{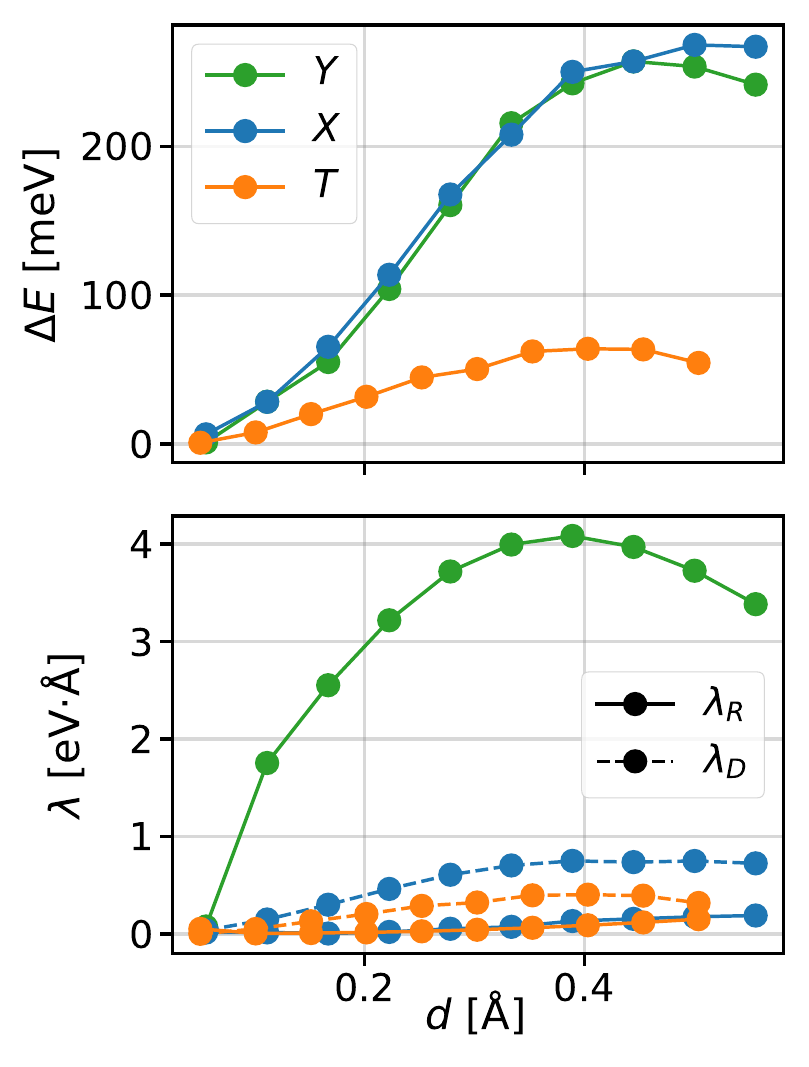}
\caption{\label{fig:n1numres} Energy splitting, $\Delta E$ (top), and corresponding Rashba and Dresselhaus parameters, $\lambda_{R/D}$ (bottom), as a functions the displacement in Cs$_{2}$PbI$_{4}$ for diagonal displacement around $X$ and $Y$, and for in-bond displacement around $T$. }
\end{figure}

In the followings, we focus on structures with $C_{2\nu}$ point symmetry but missing inversion symmetry for achieving Rashba and Dresselhaus effects. Such structures are obtained by displacing the Pb atoms uniformly in \textit{diagonal}, \textit{in-bond}, or \textit{perpendicular} direction (see Figure \ref{fig:dispdiag}, top).
Diagonal displacement leads to a face-centered orthorhombic structure with symmetry group \textit{Fmm2}, in-bond and perpendicular displacement to body-centered orthorhombic structures with symmetry group \textit{Imm2}.
The two fold $C_2$ axis is oriented along the displacement direction.
The maximum displacements were chosen 0.56 \AA\ for diagonal and 0.50 \AA\  for in-bond and perpendicular displacements, respectively. These large values are used such to highlight maximal possible effects. 

Figure \ref{fig:dispdiag} shows with the example of  Cs$_{2}$PbI$_{4}$ how these distortions influence the electronic structure. The double spin degeneracy of the lowest conduction band is lifted in some areas of the BZ in all cases, but only for diagonal and in-bond displacement this affects the band gap. In other words, Rashba effect, Dresselhaus effect, or a mixture of both only occur for diagonal and in-bond displacement. It is apparent that the splittings are considerably larger for the former. We focus on these distortion types in the following.
For diagonal displacement, the fundamental Kohn-Sham band gap of 1.79 eV is in the vicinity of the high symmetry points $Y$ and $X$ where the bands appear to be degenerate (see Brillouin zone in Figure \ref{fig:dispdiag}). Without displacement, it is located at $X$. The ${\overline{\Gamma X}}$ direction  corresponds a real-space vector that points in diagonal direction with respect to the in-plane Pb-I bonds. In case of diagonal Pb displacement, the degeneracy of states with $\mathbf{k}$ parallel and perpendicular to the displacement direction is broken.
Thus, the $X$ point splits up into the high-symmetry points $Y$ and $X$ in the new structure.
The ${\overline{\Gamma Y}}$ direction corresponds to a real-space vector that points in displacement direction, while  ${\overline{\Gamma X}}$ points perpendicular to it in the same plane. For in-bond displacement, the band gap is found in the vicinity of $T$.

\begin{figure*}[t!]
\includegraphics{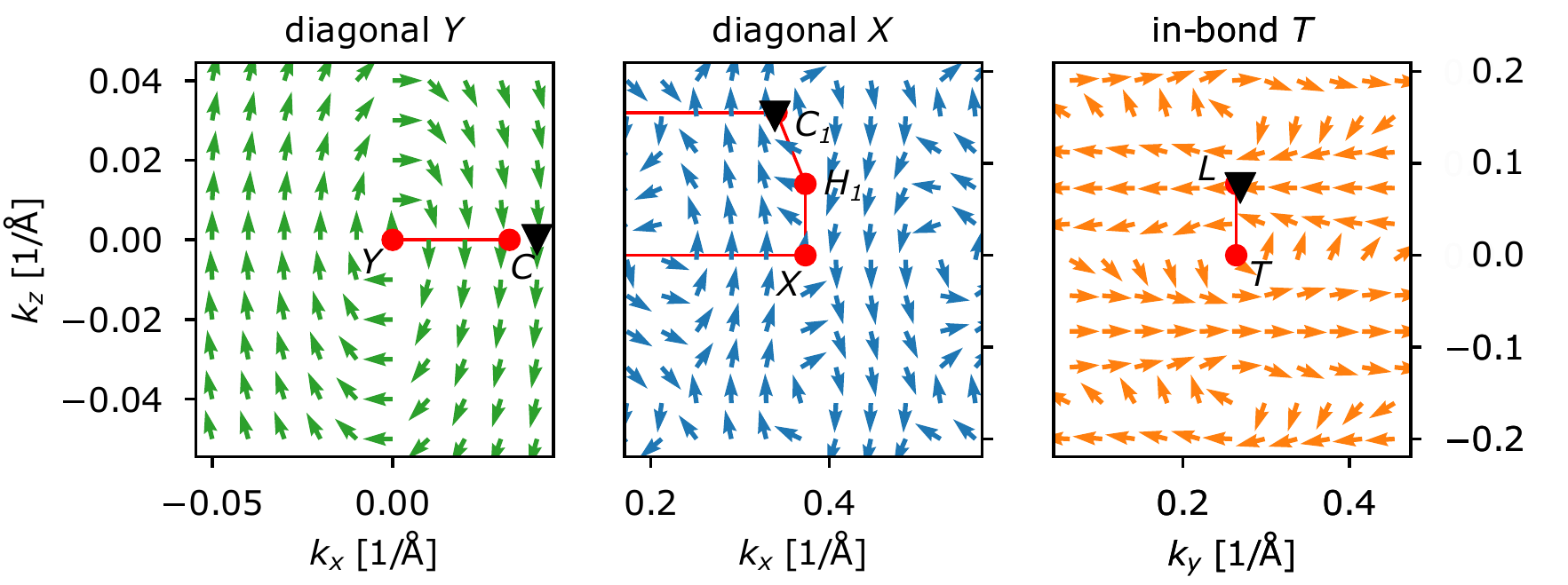}
\caption{\label{fig:spin1}Spin textures of the outer branch of the lowest conduction band of Cs$_{2}$PbI$_{4}$ for diagonal displacement ($d=0.56$Å) in the vicinity of the band gaps at $Y$ and $X$, and in-bond displacement ($d=0.50$Å). The path of the band structure in the plane is shown as red line and the high symmetry points as red dots. The black triangle indicates the position of the band gap. }
\end{figure*}

To distinguish between Rashba and Dresselhaus effect, the spin textures around the corresponding high symmetry points ($Y$, $X$ or $T$) have to be considered. 
They are  calculated in the plane perpendicular to the $C_2$ axis, \textit{i.e.}, perpendicular to the displacement direction, and are shown in Fig. \ref{fig:spin1} for diagonal displacement around $Y$ and $X$, and for in-bond displacement around $T$. 
For the former, the spin texture indicates predominantly Rashba splitting at the $Y$ point, but a mixed Rashba-Dresselhaus splitting at the $X$-point. 
Similarly, a mixed splitting around the $T$-point is observed for in-bond displacement.

In Fig. \ref{fig:n1numres}, we display the energy splitting $\Delta E$ and the parameters $\lambda_R$ and $\lambda_D$ as a function of displacement $d$. 
The latter are calculated from the spin orientation at the band gap  (Eq.~\ref{spintext}). It is apparent that $\Delta E$ is more pronounced for diagonal splitting (green and blue lines) than for in-bond displacement (orange line), as already observed in the band structure. The values at $X$ and $Y$ are very similar and only deviate for large displacements. Overall, the SO splitting exhibits a non-linear behavior.

As indicated by the spin texture, the SO splitting for diagonal displacement around $Y$ is purely Rashba like, {\it i.e.}, $\lambda_D$ is zero for all distances (see Figure \ref{fig:spin1}). Around $X$, both parameters are non-zero, but the Dresselhaus effect is dominating. For in-bond displacement, the case is similar as for diagonal displacement around $X$, but the Dresselhaus parameter is smaller. This shows that SO splitting depends on the direction of the displacement, showing up in different regions of the BZ.

For diagonal displacement, we find two basically degenerate positions of the band gap.
One is around $Y$, where the Rashba effect is decisive, and one is around $X$, caused by a mixture of both effects but clearly dominated by the Dresselhaus effect. Diagonal Pb displacement leads to much higher splitting energies and therefore is the most likely candidate to explain the strong effects found in these materials \cite{Zhai2017, Yin2018, Jansen2012, Zheng2015, Niesner2016}.

\subsection{Impact of layer thickness}
\begin{figure*}[t!]
\includegraphics{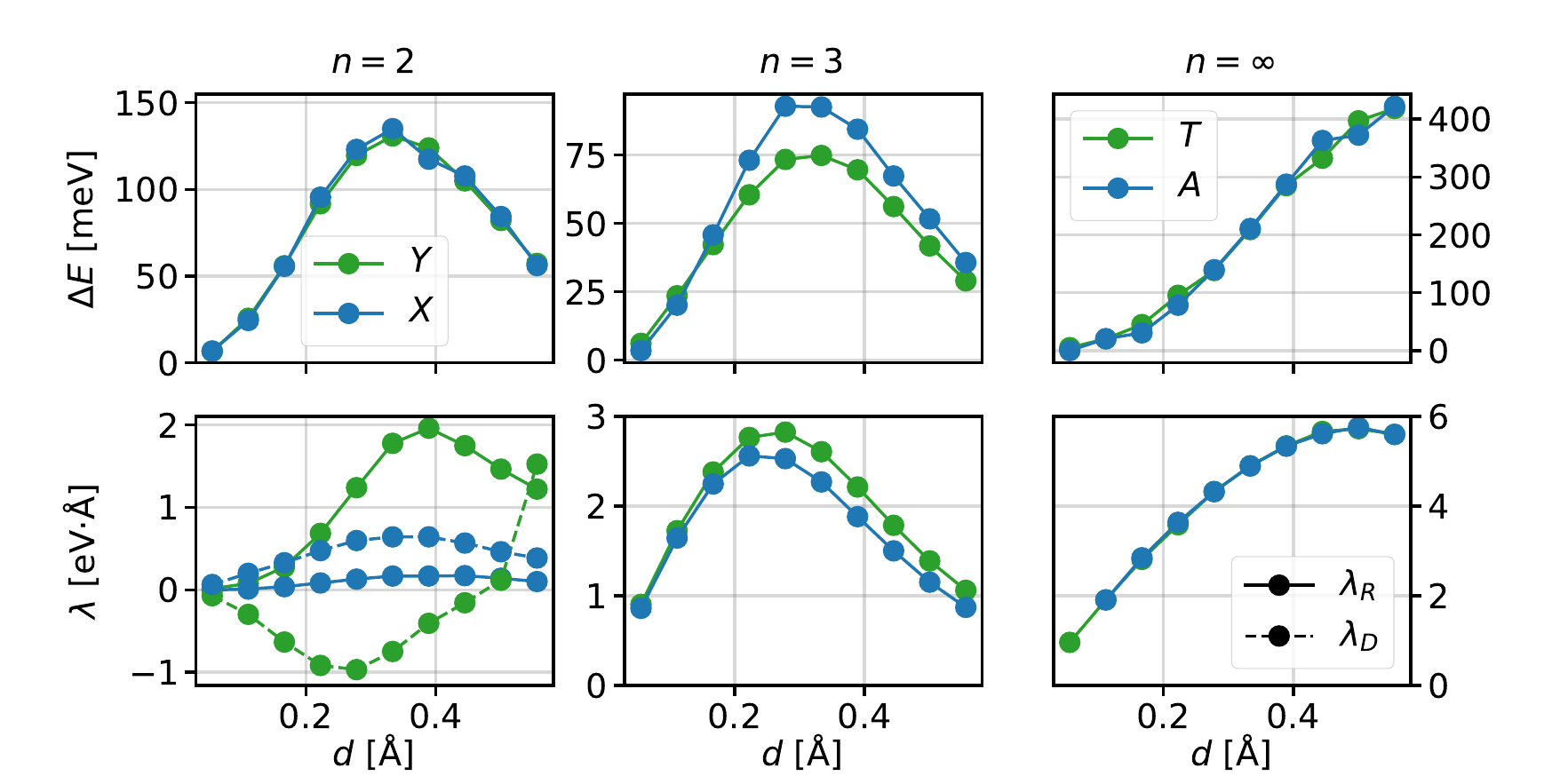}% Here is how to import EPS art
\caption{\label{fig:nlarger}{SO splittings around high symmetry points (top) and corresponding splitting parameters (bottom) as a function of displacement for $n=2$, $n=3$, and $n=\infty$.}}
\end{figure*}
\begin{figure}[t!]
\includegraphics{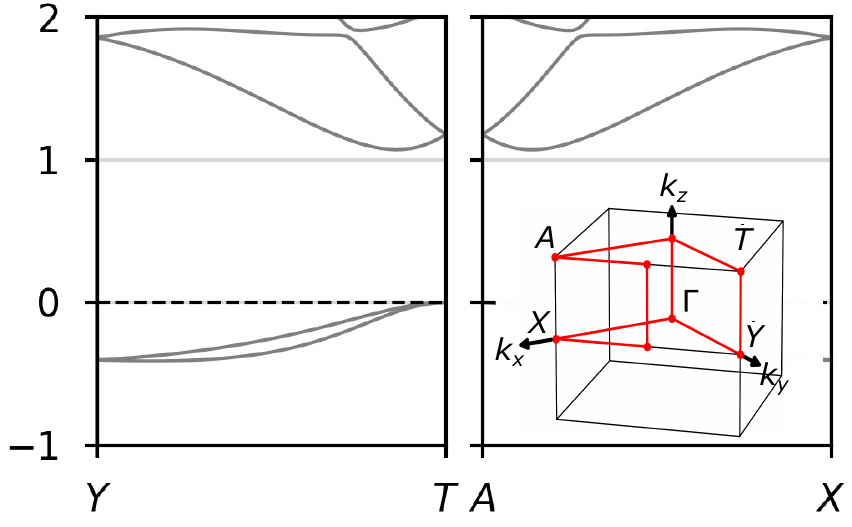}
\caption{\label{fig:bulkband} Selected paths of the band structure of CsPbI$_3$ with a diagonal Pb displacement of 0.56Å. The Brillouin zone with the high symmetry path is displayed as the small image.}
\end{figure}
\begin{figure*}[t!]
\includegraphics{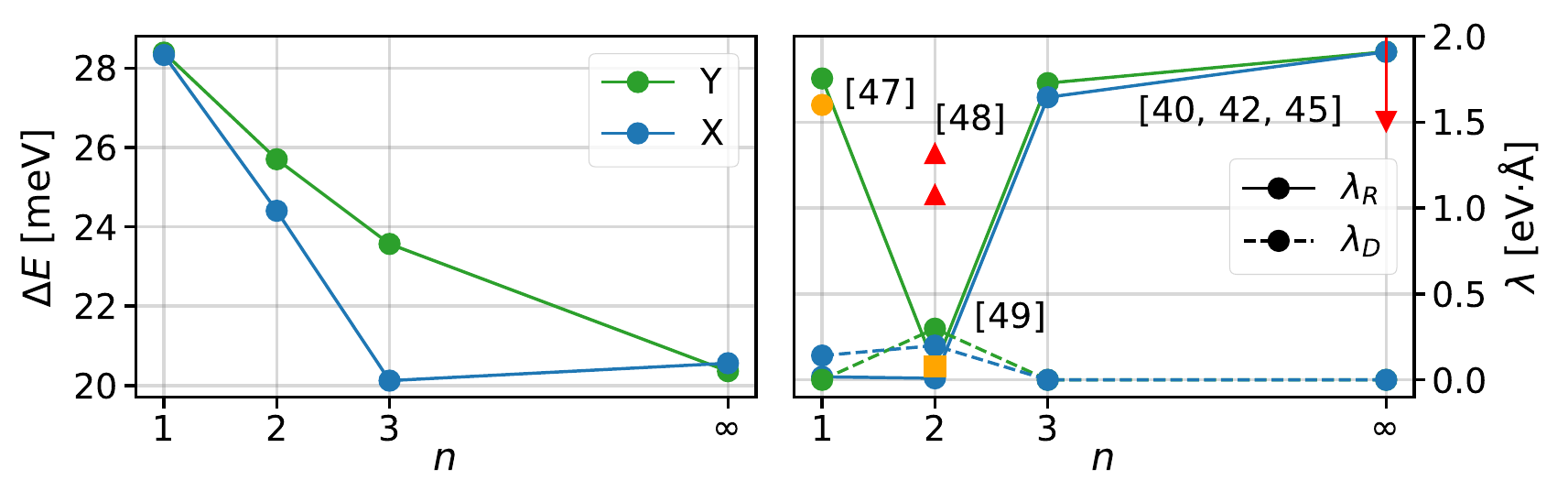}% Here is how to import EPS art
\caption{\label{fig:ncompare} {SO splitting (left) as well as Rashba and Dresselhaus parameters (right) as a function of layer thickness for a diagonal displacement of $d=0.11$ Å. Experimental (yellow) and theoretical (red) markers show reference values from literature.}}
\end{figure*}

To study the influence of increasing layer thickness on the SO splitting, we apply diagonal Pb displacement to Cs$_{n+1}$Pb$_n$I$_{3n+1}$ for $n=2, 3$, and $\infty$ and calculate $\Delta E$ and the Rashba and Dresselhaus parameters as a function of displacement. As the structures with n=1, 2, and 3 have the same symmetry, and diagonal Pb displacement corresponds to $C_{2\nu}$ point symmetry with the twofold axis oriented along the displacement direction, these materials have very similar band structures, and as before, the band gaps in the vicinity of $Y$ and $X$ are degenerate within our model. For $n=3$, they differ by about 1 meV, for $n=2$ even less. Therefore, we consider both points.

For $n=2$, the energy splittings around $Y$ and $X$ are very similar. Around $Y$, it is caused by a mixture of Rashba and Dresselhaus effect. For small displacements, the Dresselhaus effect is more pronounced, while for large displacements the Rashba effect dominates. At the maximum displacement, the situation is again reversed. Around $X$, both types of splitting occur but clearly the Dresselhaus effect is dominant for all displacements. As in the case of $n=1$, the splitting parameters are considerably larger around $Y$ than around $X$.

For $n=3$, the splitting around $X$ becomes larger than around $Y$ at decent displacements, with values smaller compared to $n=1$ and $n=2$. Around both points, it is purely caused by the Rashba effect, with a slightly larger parameter around $Y$.

The bulk structure ($n=\infty$) with diagonal displacement has higher symmetry (\textit{Imm2}) than the displaced layered structures (\textit{Fmm2}). Like the latter, the band structure shows two degenerate band gaps, close to the high-symmetry points $T$ and $A$ (note the different BZs and thus {\bf k}-paths in Figs. \ref{fig:dispdiag} and \ref{fig:bulkband}). Comparing to the cases $n=1, 2$, and 3, the $T$ point corresponds to $Y$, and $A$ to $X$, such that the $\overline{\Gamma T}$ direction corresponds to a real space vector that points in displacement direction and $\overline{\Gamma A}$ perpendicular to it. Also in this case, the energy splittings around these points show the same behavior (see Figs. \ref{fig:nlarger} and \ref{fig:bulkband}) and grow monotonously with the displacement. For large displacements, they are bigger than any splitting observed in the 2D perovskites, while for small displacements they are smaller. Around both points, $T$ and $A$, they are caused only by the Rashba effect.

So far, we have seen that diagonal Pb displacement leads to Rashba and Dresselhaus effects in 2D perovskites. Thereby, we have observed that the corresponding energy splitting is smaller for larger $n$, except for $n=\infty$. In the following, we compare our model calculations with experiments. As only small distortions have been reported in literature \cite{Zhai2017}, we choose a small displacement for the direct comparison. We find that $d=0.11$ \AA\ reproduces literature values for the splitting parameters best for all structures under investigation (see Fig.~\ref{fig:ncompare}). 

The SOC splitting around $Y$ is caused purely by the Rashba effect for all $n$, except for $n=2$ where a contribution from the Dresselhaus effect can be observed. The Rashba parameters for $n=1$ and 3 are very similar, and the one for $n=\infty$ is only slightly higher. However, for $n=2$, both parameters are significantly smaller. For $n>2$, the SO splitting is always purely caused by the Rashba effect. This suggests that pure Rashba splitting leads to much larger splitting parameters than a mixture of both effects. Interestingly, the decreasing energy splitting can not be explained with decreasing splitting parameter. The reason for this is that, with increasing $n$, the position of the band gap moves away from the reference high symmetry point, and $\Delta E \propto 1/|\mathbf{k}|$ (see Eq. \ref{splittot}).

In the experimental literature, there is to the best of our knowledge, no distinction between Rashba and Dresselhaus parameters for 2D perovskites. All of the published results are rather termed Rashba parameters. Furthermore, only results for $n=1, 2$, and $\infty$ are available. For $n=1$, Zhai and coworkers~\cite{Zhai2017} measured a Rashba parameter of about 1.6$\pm$0.1 \splitparam\ for diagonal Pb displacement in PEPI ($n=1$). Our calculated value of 1.75 \splitparam\ close to  $Y$ for Cs$_{2}$Pb$_1$I$_4$ is in good agreement with this value. In contrast, Yin \cite{Yin2018} found no Rashba effect in in the same material. As they report the structure to exhibit inversion symmetry, we assume that the displacement cannot be the same as in Ref. \cite{Zhai2017}. We note that also for PEMAPI with $n=3$, no broken inversion symmetry and thus no Rashba-Dresselhaus splitting was found in Ref. \cite{Yin2018}.

For PEMAPI with $n=2$, in Ref. \cite{Yin2018}, Rashba parameters of about 1.08 \splitparam\ and 1.32 \splitparam\ are reported for two different paths in the BZ, assigning octahedral distortion as the reason for the broken inversion symmetry. Our model for $n=2$ underestimates these parameters by far (see Fig. \ref{fig:ncompare}). On the other hand, Todd and coworkers~\cite{Todd2019} determined an experimental Rashba parameter of 0.08 \splitparam\ for bethyl-amonium-lead-iodide (BAPI; n=2), which is in very good agreement with our $\lambda_R$ around $Y$ (0.077 \splitparam). The discrepancies between Refs. \cite{Yin2018} and \cite{Todd2019} may originate from the different organic molecules separating the inorganic layers. This also explains that our model can provide qualitative insight into the underlying physics of the Rashba and Dresselhaus effect in layered perovskites but it cannot quantitatively predict the corresponding parameters for a specific material. 

For $n=\infty$, several studies have investigated SOC effects in MAPbI$_3$ \cite{Kim2014,Kepenekian2015a,Mosconi2017a}, reporting Rashba parameters in a wide range of 1.5-3.7 \splitparam. The prediction of our model, 1.91 \splitparam, lies within this range. 

\section{Conclusions}
In summary, we have introduced a model that distinguishes between Rashba and Dresselhaus effects, and establishes a clear connection between lattice distortions and SOC in the large variety of 2D hybrid perovskites. Our systematic study on symmetry breaking demonstrates that octahedral distortions in the inorganic layer lead to significant Rashba-Dresselhaus splitting. We find that mere displacement of the Pb atoms within the inorganic layers can explain the experimentally observed splittings. The largest SO splittings as well as Rashba and Dresselhaus parameters are obtained for diagonal Pb displacement, where pure Rashba splitting leads to much larger splitting parameters than a mixed Rashba-Dresselhaus effect. Pure Dresselhaus splitting does not occur for any of the studied model systems. Interestingly, we observe pure Rashba splitting only for odd $n$. For the investigated perovskites, the origin of Rashba and Dresselhaus effect can be identified as octahedral distortions corresponding to small diagonal Pb displacements. The good agreement of the computed parameters with experimental ones lead us to conclude that the Rashba-Dresselhaus effect in 2D hybrid perovksites originates largely from such Pb displacements. 

\section{Acknowledgements}
This work was partially supported by the German Research Foundation within the priority program SPP2196 “Perovskite Semiconductors,” project Nr. 424709454 and by the Deutsche Forschungsgemeinschaft (DFG) - Projektnummer 182087777 - SFB 951 (B11).

\end{document}